# Resource efficiency and Circular Economy in European SMEs: Investigating the role of green jobs and skills

Bassi, Francesca; Guidolin, Mariangela

Department of Statistical Sciences, University of Padua, Italy

**Abstract**

*Purpose of the paper*: This paper explores size and potential of green employment for Circular Economy (CE) in SMEs in the European Union and investigates the role of green jobs and skills for the implementation of CE practices.

*Design/methodology/approach*: The data are collected in a Eurobarometer survey, and refer to resource efficiency, green markets, and CE practices. Lack of environmental expertise is one of the factors that might be perceived as an obstacle when trying to implement resource efficiency actions. Previous research shows that, although resource efficiency practices are adopted by firms in all European countries, there are differences both within and between countries. The analysis of the determinants of green behavior by European SMEs is completed by a study of heterogeneity across firms and within countries with a multilevel latent class model, a hierarchical clustering method.

*Findings:* A general important evidence is that having no workers dedicated to green jobs is strongly correlated to the probability of adopting resource efficiency practices, while perceiving the need of extra environmental skills has a positive effect on the intention to implement actions in the future. Other characteristics of the firms play a significant impact on resource efficiency: in general, older and bigger firms, with larger yearly turnover, are more prone to implement actions. The type of chosen action varies across activity sectors.

*What is original/value of paper*: In the reference literature, the relationship between green jobs, environmental skills, workers employed in CE practices, and resource efficiency has received limited attention so far.

**Keywords:** resource efficiency, environmental skills, green economy, multilevel modeling,

# 1. Introduction

The idea of sustainability was born only few decades ago, due to the evidence of diminishing natural resources, climate change and increasing environmental degradation. These facts induced industries to pay attention to an appropriate management of environmental impacts of their actions. The idea of sustainability builds on the fact that there are limits to the availability of natural resources and the ability of the biosphere to absorb human activities (WCED, 1987). Sustainable development means economic growth by wasting fewer resources, conserving existing natural reserves, changing the targets of investments.

Circular Economy (CE) was also developed at the end of the last century as a promising idea for sustainable development and the first literature discussing it appeared in the 1980s. Since then, scholars and practitioners dedicated their attention to it, in order to identify principles and practices for its concrete implementation into the economic system (Lieder and Rashid, 2016; Suarez-Eiroa et al., 2019). This growing body of literature suggests many definitions of CE and, despite some criticism arising on the disorganized nature of the research concerning it (Korhonen et al., 2018), the main idea of CE refers to harmonizing economic growth and environmental protection.

CE is a widely accepted paradigm to achieve sustainability, setting limits to exploitation of resources, without rejecting economic growth (Prieto-Sandoval et al., 2018). Consequently, this approach has spread quite rapidly in many areas of the industrialized world and it is starting to be accepted also in developing countries: Abarca Guerrero et al. (2013), for example, reviewed solid waste management in urban areas of 22 developing countries, while Cecchin et al. (2019) reported a case study form Ecuador.

A popular definition of CE refers to the easy-to-remember 3Rs: reduce, reuse, and recycle (see, for example, Liu et al., 2017). Geissdoerfer et al. (2017) defined CE as "a regenerative system in which resources input and waste, emission, and energy leakage are minimized by slowing, closing, and narrowing material and energy loops. This can be achieved through long-lasting design, maintenance, repair, reuse, remanufacturing, refurbishing, and recycling", which comprehends almost all the aspects indicated above. Kirchherr et al. (2017) analyzed 114 definitions of CE and concluded that, although they combine reduce, reuse, and recycle activities, these may mean different things to different people. A number of complementary definitions of the concept emerged from the work of Lieder and Rashid (2016), emphasizing its different but important facets. In one of the first definitions provided, Stahel (1982) considered economic properties: "an economy based on a spiral-loop system that minimizes matter, energy-flow and environmental deterioration without restricting economic growth or social and technical progress". Liu et al. (2009) drawn attention to the theory and practice of the CE: the theory encompasses the principles of ecological economics that recognize that the earth's ecological system has limited resources and environmental capabilities, and, in practice, CE refers to all activities aimed at environmental protection, pollution prevention, and energy efficiency. The Ellen MacArthur Foundation (2015) defined Circular Economy as "an industrial economy that is restorative or regenerative by intention and design".

CE was introduced in 2002 by the Central Government of China to protect the environment and limit the production of pollution. In 2014, the European Commission published its 2015 Circular Economy Package with the objective of "closing the loop" of product lifecycles (European Commission, 2014 and 2015). In particular, the guidelines especially refer to product redesign in order that they are easy to maintain, repair, remanufacture or recycle (3Rs principle, Hughes, 2017). Some EU countries, for example, Finland, the Netherlands, and the UK have already adopted CE practices (Repo et al., 2018) and several studies (e.g. Stahel, 2016) showed that a shift to a Circular Economy would reduce gas emissions and increase employment. Nevertheless, implementing CE is a challenging task given the prevalence of a linear mind-set in industry and society. Circular Economy practices often require extra investments that might not be considered profitable but, as highlighted in Bocken et al. (2018), the possibility to become a sustainable economic activity requires business model innovation through experimentation, within a continuous and collective learning process with stakeholders.

The CE is strictly linked to maximization of resource efficiency. As Schulte (2013) reported, the key principles of the circular business mode are minimization of waste, understanding of the total ecosystem, flexibility of the design, use of renewable energy. These principles enhance resource efficiency. According to the reference literature, public support is an important stimulus to implement actions to improve resource efficiency (see, for example, Nussholz et al., 2018). Schulte (2013) called the attention to incentives and to taxation, suggesting to move taxes from labor to the use of non-renewable resources, in order to favor new business initiatives and obtain a positive impact on employment. Flynn et al. (2019) analyzed how different governance choices may facilitate or constrain the transition to a Circular Economy, by comparing UK and China as models of liberal and authoritarian environmental governance. Specifically, in their study, the authors highlighted the role played by the key policy instrument of standards.

As many papers in the reference literature report, SMEs are responsible for between 60 and 70% of total pollution, at least in Europe, where they represent almost 99% of the companies (European Commission, 2014). However, the knowledge of Circular Economy developed in big industries and has not spread sufficiently to small–medium enterprises (SMEs). Daddi et al. (2015) discussed the procedures to evaluate environmental impact by European SMEs and the policies to enhance green behavior. The design of efficient strategies requires a deep knowledge of the drivers of resource efficiency actions. In this context, SMEs might be seen as a reference market that needs to be segmented in order to apply targeted policies (Triguero et al., 2013). Özbuğday et al. (2020), for example, analyzing European SMEs in energy-intensive sectors, found that investing in resources efficiency has a significant effect on the growth of sales. However, as Jovè-Llopis and Segarra-Blasco (2018) showed, not all eco-strategies affect sales in the same way. One sector of activity where there is a great attention to the environment is that of accommodation and food services, at least for what concerns European firms, as it emerged from the analyses by Becken et al (2016). Colombelli et al. (2019) analyzed the impact of eco-innovation on business performance on a sample of European firms, estimating a positive effect; similar results were obtained by Leoncini et al. (2019) by a sample of Italian

SMEs. As Bleischwitz (2011) noticed, additional research is needed to assess the merits of different strategies and to identify suitable policies and levers of action. An example is Segarra-Blasco and Jovè-Llopis (2019) on the factors driving the specific policy of adopting energy efficiency and renewable energy.

Another central aspect for CE implementation is green employment. First of all, it is necessary to define CE employment in terms of skills and abilities, as for example in Burger et al. (2019). One reduced definition of green jobs counts people employed in the Environmental goods and services sector; however, this approach does not consider all those employees who use environmentally friendly processes and practices. There are various contributions in the reference literature dedicated to describe green jobs. A number of recent studies have shown that the growth of CE, among the benefits described above, produces job opportunities (Ghisellini et al., 2016). Recently, Horbach and Rammer (2019), by analyzing data collected in Germany, showed that CE innovation significantly increases employment.

However, there is still a need of investigation about employment in CE and the perceived lack of environmental competence by enterprises. Moreover, there might be sensible differences among economic activity sectors in terms of employment for CE purposes. Like CE itself, CE jobs or green jobs might be defined in several ways. The European Commission (2014) suggests that green jobs include "all jobs that depend on the environment or are created, substituted or redefined in the transition process towards a greener economy". As Garcès-Ayerbe et al. (2018) stated, empirical studies related to CE practices are scarce. A few dealt with CE drivers and barriers; however, the majority concentrate on technical and economic factors. Lack of qualified personnel is mentioned only in Kirchherr et al. (2018).

This paper explores size and potential of employment for CE in SMEs that are active in the European Union (EU) member states (MSs), and investigates the association of green jobs and green skills with CE practices in European SMEs. Previous research has suggested that although CE practices have been adopted by firms in all 28 European countries, there are differences both within countries due to firms' characteristics – size, age, turnover, type of activity – and between countries: environmental and energy-saving practices are not given the same attention everywhere in Europe (Bassi and Dias, 2019).

The enterprises considered in this study belong to all activity sectors and are a representative sample in terms of size and age. The data have been collected within the Eurobarometer surveys. The survey considered in this paper refers to resource efficiency and green markets, and collects information on actions undertaken by European firms to increase resource efficiency, thus leading to the implementation of CE practices. Lack of environmental expertise is one of the factors that might be perceived as a difficulty when trying to set up resource efficiency actions, and the number of full-time employees working in green jobs is one of the most relevant aspects to consider in this sense.

This paper is aimed at analyzing within and between countries variability with respect to implementation of resource efficiency actions and to the intention to adopt CE practices. Moreover, green employment and attitude towards green skills in European SMEs are accounted for, in order to evaluate the relationship of these factors with resource efficiency accomplishment.

This paper extends the work by Bassi and Dias (2019) in various directions. First of all, the data is more recent, collected in September 2017 vs. April 2016, on an equivalent sample of European SMEs. The data refers to a Eurobarometer survey explicitly focused on resource efficiency practices; moreover, attention is dedicated to green employment and green skills and their association with resource efficiency in European firms, a quite new topic within the reference literature. This analysis of the determinants of green behavior by European SMEs is completed by a study of heterogeneity in applying resources efficiency practices across firms and within countries with a multilevel latent class model, a hierarchical clustering method.

In this paper we evaluate which characteristics of SMEs are correlated to resource efficiency practices implementation accounting for within and between countries heterogeneity. In doing so, we test the following two hypotheses:

**Hypothesis 1**: *The number of green workers is positively correlated to resource efficiency actions implementation in European SMEs.*

**Hypothesis 2**: *The perception of lack of environmental skills is a predictor of the intention to implement resource efficiency actions in the future.*

In order to account for the nested nature of our data (SMEs within European MSs), we employ a multilevel modeling approach by considering a multilevel latent class analysis and multilevel logit models. Specifically, multilevel latent class analysis identifies groups of firms and groups of European countries that are homogeneous for their attitude towards CE practices and green employment. The multilevel approach also estimates how the homogeneous groups of companies distribute across the groups of countries: this result, combined with the fact that companies having similar attitudes towards Circular Economy may have different demographic and business profiles across groups of countries, sheds further light on the topic of green behavior and green employment in Europe. Multilevel logit models estimate the effect of the perception of the need of environmental skills on the intention to implement resource efficiency actions in the near future.

The paper is organized as follows. Section 2 briefly describes the Eurobarometer surveys, the sample and the data used for subsequent analyses, and illustrates the methodology employed. Section 3 reports the results of the analyses, along with some meaningful comments. Section 4 concludes, outlines the limitations of this paper and provides lines for future research.

## 2. Data and methodology

### 2.1 Sample characterization

Eurobarometer surveys examine European opinion and behavior on many distinct topics ranging from the support for developing countries and opinions on EU policy to the implementation of new technology, and cover citizens, households, and firms.

The data used in our study come from the Flash Eurobarometer 456, conducted in September 2017, in the EU MSs. The topic is SMEs resource efficiency and green markets. Firms with from 1 to 250 workers, operating in manufacturing, retail, industry, and services are interviewed. Questions concern actions to promote resource efficiency, green products or services, and green employment.

Table 1 reports the characterization of the sample of 13,117 SMEs operating in the 28 European MSs with reference to the number of employees, the age of the firm and the sector of economic activity, along with some selected variables referring to resource efficiency practices, green jobs and green skills. Specifically, enterprises are classified by a number of resource efficiency actions - saving water, saving energy, using renewable energy, saving materials, minimizing waste, selling scrap material, recycling, designing products easier to repair - already undertaken and planned for the subsequent two years; companies are also classified by the amount (many, some, few, no) of these actions. Information on green employment comes from the number of full-time employees working in green jobs, some or all of the time, and the expressed perception that additional green skills could improve resources efficiency. For this survey, a green job is defined as a job that directly deals with information, technologies, or materials that preserve or restore environmental quality. This requires specialized skills, knowledge, training, or experience (e.g., verifying compliance with environmental legislation, monitoring resource efficiency within the company, promoting and selling green products and services). We weighted the data in order that the sample is representative of the population of SMEs in each of the 28 countries.

Eight out of ten SMEs are very small with between one to nine employees and the great majority (80%) was founded before 2010. SMEs in the sample are active in all economic sectors, however, the majority belong to retail and manufacturing. With regard to resource efficiency practices, one third of the firms are implementing many of them, but still 10% are not undertaking any resource efficiency action. The most common resource efficiency practice being undertaken by European SMEs refers to saving inputs and to minimizing waste. Half of the companies do not employ workers in green jobs and only 20% of them perceives the lack of environmental skills. These evidences refer to the entire sample of SMEs, although there are many differences across the 28 European countries.

*Table 1. Characteristics of the SMEs in the sample*

| Size | Percentage/mean | Turnover past two years | Percentage/mean |
|---|---|---|---|
| 1-9 | 80.4 | Increased | 42.5 |
| 10-49 | 15.5 | Decreased | 21.2 |
| 50-249 | 3.0 | Unchanged | 36.3 |
| >249 | 1.1 | **Lack of expertise** | 20.1 |
| **Date of birth** | | **Full time employees in green jobs** | |
| Before 1 January 2010 | 76.9 | 0 | 57.4 |
| 1 January 2010 – 31 December 2012 | 9.3 | 1-5 | 36.0 |
| 1 January 2013 – 31 December 2017 | 12.3 | 6-9 | 3.1 |
| After 1 January 2017 | 1.5 | 10-50 | 3.0 |
| **Last year turnover in euros** | | 51-100 | 0.3 |
| <100,000 | 29.1 | >100 | 0.1 |
| 100,000-500,000 | 37.4 | **Activity sector** | 10.1 |
| 500,000-2 million | 23.2 | Manufacturing | 30.1 |
| 2-10 million | 7.2 | Retail | 43.9 |
| 10-50 million | 2.1 | Services | 15.9 |
| >50 million | 1.0 | Industry | 10.1 |
| **Actions to promote resource efficiency – present** | | **Actions to promote resource efficiency – future** | |
| Many | 33.2 | Many | 35.0 |
| Some | 31.8 | Some | 23.3 |
| Few | 24.6 | Few | 22.3 |
| No | 10.5 | No | 19.4 |
| **Undertaken actions** | | **Future actions** | |
| Minimising waste | 65.5 | Saving energy | 58.8 |
| Saving energy | 63,2 | Minimasing waste | 56.6 |
| Saving materials | 56.8 | Saving materials | 50.8 |
| Saving water | 47.2 | Saving water | 45.1 |
| Recycling or reusing waste | 41.8 | Recycling or reusing waste | 37.9 |
| Designig substainable products | 25.3 | Designig substainable products | 24.6 |
| Selling scrap material | 21.1 | Using renewable energy | 22.4 |
| Using renewable energy | 14.0 | Selling scrap material | 21.0 |

Table 2 describes the sample at the country level, providing insights into the variability between countries in terms of percentage of firms undertaking the surveyed resource efficiency actions, declaring the need for environmental expertise in order to improve resource efficiency, and the number of full-time employees dedicated to green jobs. These figures are comparable since statistics are calculated with weighted data that account for the differences in the number of firms in the various countries. These three variables have significantly different distributions across the 28 EU MSs. Estonia is the country with, on average, less employees in green jobs: this is also the country where SMEs do not acknowledge the need of specific environmental expertise. Ireland has the highest percentage of firms with more than six green-job employees; in Spain, we observe the highest percentage of firms calling for additional environmental skills. Spain, United Kingdom, Ireland, and Portugal show high percentages of SMEs implementing many resource efficiency actions; on the opposite side, we find Bulgaria, Estonia, Lithuania and Romania.

*Table 2. EU countries by the percentage of firms undertaking resource efficiency practices, lacking environmental expertise, and number of employees with green jobs*

| Country | Lacking environmental expertise | Undertaken actions | | | | Employees in green jobs | | |
|---|---|---|---|---|---|---|---|---|
| | | Many | Some | Few | No | 0 | 1-5 | 6+ |
| | | Percentages | | | | | | |
| Austria | 18.1 | 38.1 | 32.0 | 22.2 | 7.7 | 55.2 | 35.1 | 9.7 |
| Belgium | 16.9 | 38.7 | 30.8 | 21.5 | 9.1 | 64.8 | 31.2 | 4.0 |
| Bulgaria | 11.1 | 9.4 | 23.2 | 33.1 | 34.3 | 73.5 | 20.0 | 6.5 |
| Republic of Cyprus | 9.5 | 14.8 | 22.2 | 40.7 | 22.2 | 81.5 | 11.1 | 7.8 |
| Czech Republic | 21.8 | 30.5 | 34.7 | 21.2 | 13.5 | 77.2 | 18.7 | 4.1 |
| Germany | 14.9 | 30.6 | 36.1 | 25.9 | 7.4 | 58.1 | 32.1 | 9.8 |
| Denmark | 13.2 | 27.2 | 29.6 | 28.0 | 15.2 | 67.7 | 24.4 | 7.9 |
| Estonia | 0 | 2.9 | 5.9 | 38.2 | 52.9 | 85.3 | 14.7 | 0 |
| Spain | 31.6 | 45.0 | 27.3 | 21.9 | 5.7 | 48.0 | 43.2 | 8.8 |
| Finland | 13.9 | 23.5 | 28.8 | 31.1 | 16.7 | 44.9 | 51.2 | 4.8 |
| France | 33.7 | 40.9 | 38.1 | 15.4 | 5.6 | 68.9 | 26.3 | 4.8 |
| United Kingdom | 15.4 | 46.0 | 32.5 | 16.9 | 4.6 | 70.6 | 25.6 | 3.8 |
| Greece | 22.9 | 20.6 | 24.4 | 28.6 | 26.4 | 51.9 | 35.4 | 12.7 |
| Croatia | 13.2 | 26.2 | 39.3 | 25.0 | 9.5 | 36.4 | 58.4 | 5.2 |
| Hungary | 11.0 | 16.3 | 27.6 | 38.2 | 18.0 | 76.8 | 16.9 | 6.3 |
| Ireland | 24.0 | 44.4 | 37.0 | 13.0 | 5.6 | 56.5 | 30.4 | 13.1 |
| Italy | 12.1 | 29.0 | 29.9 | 31.5 | 9.7 | 42.7 | 51.8 | 5.5 |
| Lithuania | 7.1 | 5.9 | 23.5 | 37.6 | 32.9 | 81.0 | 14.3 | 4.7 |
| Luxemburg | 17.6 | 30.0 | 25.0 | 30.0 | 15.0 | 57.9 | 36.8 | 5.3 |
| Latvia | 14.6 | 17.0 | 32.1 | 28.3 | 22.6 | 30.0 | 58.0 | 12.0 |
| Malta | 7.7 | 21.4 | 35.7 | 35.7 | 7.1 | 92.9 | 7.1 | 0 |
| The Netherlands | 13.1 | 29.6 | 37.9 | 25.1 | 7.4 | 63.0 | 30.6 | 6.4 |
| Poland | 27.7 | 24.3 | 31.8 | 27.3 | 16.7 | 58.5 | 34.7 | 6.8 |
| Portugal | 17.2 | 48.6 | 28.5 | 18.3 | 4.6 | 48.4 | 46.0 | 5.6 |
| Romania | 17.3 | 8.5 | 17.7 | 37.9 | 35.9 | 62.3 | 28.5 | 9.2 |
| Sweden | 13.9 | 41.3 | 35.4 | 16.7 | 6.6 | 40.1 | 51.4 | 8.5 |
| Slovenia | 17.5 | 26.0 | 27.4 | 24.7 | 21.9 | 72.9 | 25.7 | 1.4 |
| Slovakia | 7.9 | 21.8 | 27.7 | 36.8 | 13.6 | 38.8 | 52.6 | 8.6 |

## 2.2. Methodology: Multilevel Modeling

### 2.2.1. The multilevel latent class model

Latent class (LC) analysis provides models that consider explicitly the effect of one or more latent variables when studying relationships between observed variables, and accounts for the categorical nature of these variables.

Latent class models were introduced by Lazarsfeld and Henry (1968) to express latent attitudinal variables from dichotomous survey items, then they were extended to nominal variables by Goodman (1974a, 1974b), who also developed the maximum likelihood algorithm for estimating latent class models that serves as the basis for many software with this purpose. Later, these models were further extended to include observable variables of mixed scale type, like ordinal, continuous and counts.

Let:

$Y_{ijk}$, $i=1,...,I$, $j=1,...J$, $k=1,...K$, denote the response of individual or level-1 unit $i$ within group or level-2 unit $j$ on indicator or item $k$;

$s_k = 1,...S_k$, a particular level of item $k$;

$Z_{ij}$, a latent variable with $L$ classes;

$l$, a particular latent class, $l=1,...,L$;

$\underline{Y}_{ij}$, the full vector of responses of case $i$ in group $j$;

$\underline{s}$, a possible response pattern.

The probability structure defining a simple LC model may be expressed as follows:

$$P(\underline{Y}_{ij} = \underline{s}) = \sum_{l=1}^{L} P(Z_{ij} = l) P(\underline{Y}_{ij} = \underline{s} \mid Z_{ij} = l) = \sum_{l=1}^{L} P(Z_{ij} = l) \prod_{k=1}^{K} P(Y_{ijk} = s_k \mid Z_{ij} = l) \quad (1)$$

As specified in equation (1), the probability of observing a particular response pattern is a weighted average of class-specific probability $P(Y_{ijk} = s_k \mid Z_{ij} = l)$, weight being the probability that unit $i$ in group $j$ belongs to latent class $l$. As the local independence assumption implies, indicators $Y_{ijk}$ are assumed to be independent conditional on LC membership. This model is also referred to as traditional LC cluster model within the relevant literature.

Multilevel latent class modelling (Vermunt, 2003) is an approach based on the assumption that some model parameters can vary across groups, clusters or level-2 units. As an example of hierarchical data, operations are nested in a bank's customers: operations are level-1 units, clients are level-2 units (Bassi, 2017). This is different from traditional latent class modelling, which assumes that the parameters are the same for the whole sample. The multilevel approach allows for variation across level-2 units for the intercept (threshold) of each latent class indicator. This makes it possible to examine how level-2 units influence the level-1 indicators that define latent class membership. This method adopts a random-effects approach rather than a fixed-effects approach, enabling the effects of level-2 covariates to be verified on the probability of belonging to a given latent class.

A multilevel LC model (Vermunt, 2003) consists of a mixture model equation for level-1 and level-2 units, in which a group-level discrete latent variable is introduced so that the parameters are allowed to differ across latent classes of groups:

$$P(\underline{Y}_{ij} = \underline{s}) = \sum_{h=1}^{H} \left[ P(W_j = h) \prod_{i=1}^{n_j} \left[ \sum_{l=1}^{L} P(X_{ij} = l | W_j = h) \prod_{k=1}^{K} P(Y_{ijk} = s_k | Z_{ij} = l) \right] \right] \quad (2)$$

where

$W_j$ denotes the latent variable at the group level, assuming value $h$, with $h=1,...,H$;

$n_j$ is the size of group $j$.

Equation (2) is obtained with the additional assumption that $n_j$ members' responses are independent of one another conditional on group class membership.

A natural extension of the multilevel LC model involves including level-1 and level-2 covariates to predict membership, like an extension of the LC model with concomitant variables (Dayton & McReady, 1988).

In the terminology of MLLC modelling, the categories of the latent variable for level-1 units are called clusters, while the categories of the latent variable for level-2 units are called classes. In our application, level-1 units are companies and level-2 units are European countries.

### 2.2.3. The multilevel logit regression model

Multilevel regression models (Hedeker and Gibbons, 1994; Hox, 2002; Snijders and Bosker, 2012) estimate simultaneously at two levels. Level-1 units are represented by firms, level-2 units by countries. Using the same notation introduced in the preceding paragraph, a multilevel regression model is specified by:

$$y_{ij} = \beta x_{ij} + u_j + \varepsilon_{ij}$$

where $x_{ij}$ is the vector containing the values of the covariates for observation $i$ in cluster $j$, $\beta$ is the vector of parameters (fixed effects), $u_j$ is the random effect for group $j$ - this random effect represents factors affecting $y_{ij}$ that are shared within class $j$, after controlling for individual covariates -, and $\varepsilon_{ij}$ is the error term with the usual assumptions: errors are independently distributed as Normal with 0 mean and equal variance.

### 3. Results
### 3.1 Multilevel Latent Class Model

In this section, we report the results of two multilevel LC analyses that have been conducted on the data[1]. These involve *undertaken* and *future* actions of resource efficiency across SMEs in the EU countries.

---

[1] Model estimation was performed with Latent Gold (Vermunt and Magidson, 2013)

These analyses are aimed at testing our Hypothesis 1:

**H1**: *The number of green workers is positively correlated to resource efficiency actions implementation in European SMEs*.

In order to perform the analysis, we also consider other SMEs characteristics as age, size, sector of economic activity, that might explain both between countries and within country variability.

### 3.1.1 Resource efficiency actions across SMEs and EU countries: *undertaken actions*

The first model aims at identifying groups of SMEs that are homogeneous for the amount and type of resource efficiency actions already undertaken. Indicators are the eight binary variables describing the resource efficiency actions proposed in the questionnaire. The best fitting multilevel latent class (MLLC) model identifies six clusters of firms and four classes of countries. This is the model that shows the lowest BIC index, comparing specifications with different combinations of the number of clusters and classes. The statistically significant variables are SME's turnover, age, sector of economic activity and the number of full-time employees dedicated to green jobs.

*Table 3. MLLC model — estimation results - cluster profiles for undertaken actions*

|  | Cluster 1 | Cluster 2 | Cluster 3 | Cluster 4 | Cluster 5 | Cluster 6 |
|---|---|---|---|---|---|---|
|  | *All actions* | *All actions except renewable energy* | *Saving actions* | *Reducing waste actions* | *Renewable energy action* | *No actions* |
| Size | 13.74 | 14.23 | 20.39 | 14.76 | 7.23 | 29.63 |
|  | **Conditional probabilities %** | | | | | |
| Class 1 | 5.57 | 8.72 | **23.91** | 8.56 | 1.53 | **51.72** |
| Class 2 | 9.85 | 14.59 | **20.87** | **23.04** | 4.50 | **27.15** |
| Class 3 | **33.61** | 1.96 | 16.24 | 7.25 | **26.10** | 14.84 |
| Class 4 | 11.50 | **37.39** | 18.19 | **21.74** | 0.00 | 11.17 |
| *Indicators* | | | | | | |
| **Saving water** | **58.71** | **91.10** | **78.16** | 26.35 | 9.43 | 2.86 |
| **Saving energy** | **87.88** | **94.96** | **99.10** | 30.33 | 59.46 | 14.15 |
| **Using renewable energy** | 44.30 | 9.25 | 11.15 | 0.00 | **39.58** | 2.58 |
| **Saving materials** | **87.95** | **90.56** | **63.02** | 50.98 | 43.55 | 9.5 |
| **Minimasing waste** | **91.70** | **93.38** | **74.54** | **75.36** | 48.66 | 12.74 |
| **Selling scrap material** | **53.18** | 28.83 | 10.66 | 18.55 | 12.77 | 6.69 |
| **Recycling or reusing waste** | **70.05** | **71.66** | 27.44 | **44.41** | 28.54 | 11.36 |
| **Designig substainable products** | 46.88 | **51.23** | 12.23 | 22.18 | 17.54 | 4.75 |
| *Covariates* | | | | | | |
| **Activity sector** | | | | | | |
| Manufacturing | **14.33** | **11.95** | 6.73 | **11.85** | 8.07 | 7.83 |
| Retail | 27.98 | **34.41** | **34.16** | 26.85 | 18.51 | 30.80 |
| Services | 29.12 | 43.77 | **48.33** | 40.31 | **48.45** | **51.31** |
| Industry | **28.56** | 9.87 | 10.78 | **21.00** | **24.97** | 10.07 |
| **Date of birth** | | | | | | |
| Before 1 January 2010 | **82.06** | 75.80 | **78.30** | 74.40 | 75.95 | 73.06 |
| 1 January 2010 – 31 December 2012 | 7.39 | **10.15** | 7.07 | 9.68 | 8.07 | **11.82** |
| 1 January 2013 – 31 December 2016 | 10.30 | 11.37 | 12.21 | **13.31** | 11.51 | **13.74** |
| After 1 January 2017 | 0.14 | **1.84** | **1.77** | 2.03 | 3.43 | 0.77 |
| **Full time employ green jobs** | | | | | | |
| 0 | 27.58 | 39.19 | **58.01** | 63.90 | 62.33 | **70.25** |
| 1-5 | **53.48** | 46.54 | 30.38 | 29.36 | 29.59 | 17.31 |
| 6-9 | **5.25** | **3.31** | **3.58** | 1.33 | 1.94 | 2.16 |
| 10-50 | **6.61** | **5.52** | 1.98 | 0.07 | 1.90 | 1.11 |
| 51-100 | 0.00 | 0.76 | **0.37** | 0.02 | 0.12 | 0.03 |
| >100 | **0.78** | 0.00 | 0.00 | 0.00 | 0.00 | 0.00 |
| **Last year turnover in euros** | | | | | | |
| <100,000 | 20.83 | 18.01 | **27.45** | 22.85 | 20.53 | **32.34** |
| 100,000-500,000 | 26.08 | **33.23** | **36.00** | 30.79 | 26.27 | **31.72** |
| 500,000-2 million | **27.19** | 22.24 | 11.60 | 18.35 | **32.86** | 15.08 |
| 2-10 million | 9.99 | 4.97 | 5.58 | **6.47** | **8.42** | 3.71 |
| 10-50 million | 8.35 | 0.06 | 0.18 | 0.41 | **4.70** | 0.28 |
| >50 million | 0.21 | **2.30** | **1.03** | 0.23 | 0.00 | 0.61 |

Clusters in Table 3 are ordered by the number of resource efficiency actions already implemented by the group of SMEs. Percentages in bold indicate figures greater than the sample mean and aim at favoring interpretation. In cluster 1, almost 14% of SMEs are classified. These firms undertake all eight resource efficiency actions considered in the Eurobarometer survey. SMEs in cluster 2 (14%) are already undertaking all actions, except the use of predominantly renewable energy. SMEs classified in cluster 3, that accounts for 21% of the sample, are implementing practices for saving production inputs, water, energy, materials and for minimizing waste, which is an action strictly related to savings. In cluster 4, we find a 14% of firms that are mainly concentrated in reducing waste. In cluster 5 (7%), SMEs adopt the only policy of using predominantly renewable energy and finally, cluster 6, the largest one, contains those firms that are not implementing any of the proposed green actions. The statically significant covariates indicate that some characteristics of the European SMEs have an impact on their attitude towards the implementation of those CE practices. Specifically, for example, the fact of being established before 2010, having a yearly turnover greater than 500,000 Euros, having employees devoted to green jobs, and operating in the sectors of manufacturing and industry, have a positive impact on belonging to cluster 1. On the contrary, SMEs classified in cluster 6 were established after 2010, do not have employees for green jobs, yearly turnover is below 500,000 Euro and operate mainly in the sector of services. As a general evidence, number of employees for green jobs, size and age of the firm have a positive impact on the number of resource efficiency actions undertaken as well as operating in the sector of manufacturing and industry.

The upper part of Table 3 shows how the six clusters of SMEs are distributed across homogeneous groups of European countries. The four homogeneous classes of European countries are composed as follows. Class 1 is composed by Bulgaria, Cyprus, Estonia, Greece, Hungary, Latvia, Lithuania, Romania, Slovakia. Class 2 is composed by Czech Republic, Denmark, Finland, Croatia, Italy, Luxemburg, Poland, and Slovenia. In Class 3 we find Austria, Belgium, Germany, Malta, The Netherlands, and Sweden. In class 4, Spain, France, United Kingdom, Ireland and Portugal. It is interesting to note that these four groups of countries, that are homogenous for the type of firms operating on their territory, do not coincide with the usual geographical classification of EU MSs in the four areas: North, South, East and West, used at international statistical level. The conditional probabilities reported in Table 3 show that countries in class 1 are associated with SMEs of the type in cluster 6 (no actions); countries in class 2 are associated with firms in clusters 3 (saving), 4 (waste), and 6 (no actions); this is the group of countries with the largest amount of heterogeneity in the type of firms. In countries of class 3, we find SMEs that pay the greatest attention to resources efficiency, those of cluster 1 (all actions), however there is also a non-negligible percentage of firms classified in cluster 5 (renewable energy). Lastly, in class 4, SMEs belong to clusters 2 (all actions except use of renewable energy) and 4 (waste). However, in all classes of countries there are non-negligible percentages of SMEs belonging to all clusters, except for cluster 5 in class 4, indicating an important level of heterogeneity within countries with reference to the attitude of SMEs towards resources efficiency and CE in general.

**3.1.2 Resource efficiency actions across SMEs and EU countries:** *future actions*

The second multilevel latent class model identifies groups of SMEs on the basis of their intention to implement the eight resource efficiency actions in the next two years. Again, the best fitting model, in terms of BIC index, has six clusters and four classes, but in this case two new variables resulted statistically significant in forming the clusters: a binary variable indicating if the firm recognizes the lack of specific environmental skills in order to set up resource efficiency actions and the change in turnover over the past two years. Results are reported in Table 4.

*Table 4. MLLC model — estimation results - cluster profiles for future actions*

|  | Cluster 1 | Cluster 2 | Cluster 3 | Cluster 4 | Cluster 5 | Cluster 6 |
|---|---|---|---|---|---|---|
|  | *All actions* | *All actions except renewable energy* | *Saving and waste reduction actions* | *Some actions* | *Waste reduction actions* | *No actions* |
| Size | 16.81 | 8.22 | 16.77 | 6.53 | 10.22 | 41.46 |
|  | **Conditional probabilities %** | | | | | |
| Class 1 | 12.55 | 4.01 | 9.77 | 6.08 | 7.14 | **60.44** |
| Class 2 | 14.87 | 7.32 | **27.17** | 1.71 | **15.09** | **33.84** |
| Class 3 | 15.52 | 4.92 | 12.90 | **24.82** | 9.25 | **32.59** |
| Class 4 | **33.46** | **24.51** | 15.79 | 0.01 | 8.41 | 17.82 |
| *Indicators* | | | | | | |
| **Saving water** | 86.17 | 85.74 | 77.34 | 27.00 | 4.22 | 2.06 |
| **Saving energy** | 98.35 | 92.47 | 88.59 | 92.73 | 14.32 | 11.76 |
| **Using renewable energy** | **48.79** | 7.32 | **25.95** | **63.26** | 3.59 | 8.43 |
| **Saving materials** | **98.43** | **73.44** | **68.76** | 46.89 | 47.41 | 3.96 |
| **Minimasing waste** | **97.20** | **90.89** | **78.17** | 52.49 | **69.15** | 1.97 |
| **Selling scrap material** | **56.09** | **31.37** | 7.50 | **27.84** | 17.71 | 2.83 |
| **Recycling or reusing waste** | **88.90** | **52.16** | 29.61 | 25.55 | **42.47** | 6.61 |
| **Designig substainable products** | **68.52** | **28.21** | 15.98 | **27.27** | 21.60 | 1.79 |
| *Covariates* | | | | | | |
| **Activity sector** | | | | | | |
| Manufacturing | **13.36** | 0.09 | 7.73 | **13.24** | **11.66** | 8.73 |
| Retail | 30.37 | 30.88 | **32.07** | **35.41** | 27.96 | 28.17 |
| Services | 38.57 | 38.59 | **49.64** | 42.94 | 36.80 | **48.48** |
| Industry | **17.70** | **21.52** | 10.56 | 8.41 | **23.58** | 14.62 |
| **Date of birth** | | | | | | |
| Before 1 January 2010 | 71.32 | **84.07** | 74.68 | **79.84** | **78.24** | 76.03 |
| 1 January 2010 – 31 December 2012 | **10.24** | 5.51 | **10.46** | 7.53 | 7.32 | **10.51** |
| 1 January 2013 – 31 December 2016 | **16.57** | 7.11 | 12.09 | **12.61** | 11.65 | 11.86 |
| After 1 January 2017 | 1.46 | 3.24 | **1.60** | 0.00 | **1.72** | 0.97 |
| **Full time employees in green jobs** | | | | | | |
| 0 | 26.41 | **74.11** | 53.56 | 43.64 | 53.96 | **64.90** |
| 1-5 | **56.91** | 18.43 | **34.02** | 36.10 | **38.99** | 23.53 |
| 6-9 | **3.71** | 2.25 | **3.72** | 8.79 | 0.86 | 1.85 |
| 10-50 | **7.33** | 1.40 | 1.36 | **4.62** | 0.51 | 1.88 |
| 51-100 | **0.30** | 0.00 | **0.04** | 0.95 | **0.47** | 0.18 |
| >100 | 0.00 | 0.74 | 0.04 | 0.00 | 0.00 | 0.04 |
| **Last year turnover in euros** | | | | | | |
| <100,000 | 21.00 | 15.70 | **31.46** | 12.48 | 21-31 | **29.36** |
| 100,000-500,000 | 26.91 | **32.56** | 35.78 | 24.12 | 30.59 | **33.00** |
| 500,000-2 million | **24.34** | **29.25** | 7.52 | **37.36** | 22.39 | 14.85 |
| 2-10 million | **7.90** | 6.23 | 4.89 | **11.18** | 4.29 | 5.15 |
| 10-50 million | **3.61** | 1.35 | 0.10 | **9.17** | 0.73 | 0.63 |
| >50 million | 0.34 | 0.38 | **2.26** | 0.00 | 1.30 | 0.53 |
| **Turnover over the past two years** | | | | | | |
| Increased | **45.04** | 34.80 | 38.91 | **52.03** | 36.74 | 37.25 |
| Decreased | 14.34 | **28.79** | 17.71 | 19.17 | 17.28 | **21.50** |
| Unchanged | **34.09** | 29.52 | 33.35 | 27.35 | **39.62** | 35.06 |
| **Need of expertise** | **22.66** | **61.60** | 0.00 | **22.12** | 14.82 | 7.51 |

Clusters in Table 4 are ordered by the number of resource efficiency actions that the group of SMEs plan to implement in the next resource two years. Percentages in bold indicate figures greater than the sample mean, in order to aid interpretation. Nearly 17% of SMEs are classified into cluster 1. These firms intend to implement all eight resource efficiency actions considered in the Eurobarometer survey. SMEs in cluster 2 (8%) are planning all actions except the use of predominantly renewable energy. In cluster 3, we find a 17% of firms that are mainly concentrated in input saving and waste reduction for the near future. SMEs classified in cluster 4, that represents 6% of the sample, aim at introducing practices with regard to the use of energy, product and scrap material. In cluster 5 (10%), SMEs will adopt policies concerning waste and, finally, cluster 6 contains those firms that will not implement any of the proposed green actions, this is the largest group (41%). As observed before, the number of employees for green jobs, the dimension and the age of the firm have a positive impact on the number of resource efficiency actions in program, as well as operating in the sectors of manufacturing and industry. Perceiving the lack of green skills has also a positive effect on belonging to clusters 1, 2 and 4 where more resource efficiency actions are on the way to be implemented. Specifically, in cluster 2 we find SMEs, for the majority, with no employees in green jobs but with a clear perception that this type of skills would improve resource efficiency.

The upper part of Table 4 shows how the six clusters of SMEs are distributed across homogeneous groups of European countries. The four homogeneous classes of European countries are composed as follows. Class 1 is composed by Bulgaria, Cyprus, Denmark, Estonia, Finland, Greece, Hungary, Luxemburg, Malta, Portugal, Sweden, and Slovenia. Class 2 is composed by Czech Republic, Croatia, Hungary, Italy, Latvia, Lithuania, Poland, Romania, and Slovakia. In class 3 we find Austria, Belgium, Germany, and The Netherlands. In class 4, Spain, France, The United Kingdom and Ireland. The conditional probabilities reported in Table 5 show that countries in class 1 are associated with SMEs of the type in cluster 6 (no actions); countries in class 2 are associated with firms in clusters 3 (saving and waste), 5 (waste), and 6 (no actions). In class 3, SMEs belong to clusters 4 (energy, product, scrap material) and 6 (no actions). In countries of class 4, we find SMEs that pay the greatest attention to resources efficiency, those of clusters 1 (all actions), and 2 (all actions except the use of renewable energy). However, in all classes of countries there are non-negligible percentages of SMEs belonging to all clusters, indicating an important level of heterogeneity within countries with reference to the attitude of SMEs towards resources efficiency and CE in general.

The four classes of countries show some differences in their composition in the two analyses (for present and future actions). For example, class 1 in the second MLLC model appears as a combination of two groups of countries: those where SMEs, in a great percentage, do not implement nor intend to introduce CE practices (Bulgaria, Cyprus, Estonia, Greece), and those where many SMEs are already active on the environmental side but do not intend to add other resource efficiency actions (Slovenia, Denmark, Finland, Luxemburg, Malta, Sweden). A deeper insight into the data reports that correlation coefficients among each couple of variables referring to the same action, undertaken and planned, are not so high, lower than 0.6, and that there are important percentages of SMEs that intend to implement a new action. These percentages range from 8.2 with reference to selling scrap material to another company to 27.5 for saving energy. These proportions, moreover, vary across the 28 European countries and are significantly associated to the number of full-time employees working in green jobs and the perception of lack of specific environmental expertise. A cluster analysis of countries using as input the above described variables identifies three homogeneous groups. In the first cluster, we find Bulgaria, Cyprus, Denmark, Estonia, Hungary, Latvia, Malta, and Slovenia, countries with the lowest percentage of SMEs that plan to implement an additional resource efficiency action. The smallest group, Spain, France, United Kingdom, and Ireland, has the highest percentages of intention to adopt new CE practices. A third cluster, Austria, Belgium, Czech Republic, Germany, Finland, Greece, Hungary, Italy, Luxemburg, Latvia, The Netherlands, Poland, Portugal, Romania, Sweden, and Slovakia, with an intermediate proportion.

The above results suggest that the number of employees dedicated to green jobs and the perception of need of green skills among the labor force are associated to the decision to introduce new resource efficiency actions by European SMEs. The employment of these skills could be encouraged in order to promote CE adoption. This could be part of green policies especially in those European countries where the composition of SMEs is more in favor of the type not devoted to green actions.

### 3.2 Multilevel Regression models: green jobs and skills for resource efficiency

In this section we present the estimates of nine multilevel regression models, to test our Hypothesis 2

**H2**: *The perception of lack of environmental skills is a predictor of the intention to implement resource efficiency actions in the future.*

In particular, we aim at investigating which factors might affect the decision to implement in the near future resource efficiency actions, with a special attention to the number of workers employed in green works and the perceived need of environmental skills.

Tables 5 provides the estimates of the proposed multilevel regression models. In the first column the dependent variable is ordinal, indicating the amount of actions (no, few, some, many) that SMEs are planning to implement as additional resource efficiency initiatives in the next two years. In the other eight models, the dependent variable assumes a value of 1 if the firm will implement the resource efficiency action reported in each column. The independent variables characterizing the firms are: the number of employees working in green jobs, the age of the firm, the total turnover in 2016, the sector of economic activity, the change in turnover over the past two years, and the perception of lack of environmental skills. In all models, a random effect at the country level is specified in order to account for differences across countries. The coefficients reported are estimated with effect coding, i.e., the parameters referring to one covariate sum up to 0 so that the category-specific effects should be interpreted in terms of deviation from the average. The intra-class correlation coefficient (ICC) is the proportion of the total dispersion that is explained by the country level. The multilevel regression model estimates the effect of firms' characteristics on resource efficiency behavior, considering the fact that SMEs operating in the same country may show similarities.

Results show that the age of the SMEs has no effect on undertaking future actions to increase resource efficiency, and the same result appears also in the multilevel logit regression model for all eight single actions. Other interesting results in common with the eight logit models are that not having full-time employees dedicated to green jobs decreases the number of resource efficiency actions for the future and a positive effect on the independent variable is estimated for the fact that the SME declares the lack of specific environmental skills. Yearly turnover greater than 10 million euros has a positive effect, turnover lower than 500,000 euros has a negative impact. Change in the turnover has a significant impact on the number of actions that the SME is planning for the future, this result is replicated also for the single actions except saving water. The fact that the firm operates in the manufacturing sector increases the intention to implement actions, the contrary is estimated for retail and services. Because the variance of the random effect is positive and significantly different from 0, there is heterogeneity in this behavior between countries. The intra-class correlation coefficient (ICC) is 0.0540, i.e., the country level accounts for 5.12% of the variability. As a general remark, the result confirms the evidence reported in the previous analysis.

### 3.2.1 Saving water

An average of 47.2% of firms in Europe implement saving water policies and 54.9% are planning to implement it in the future as an additional policy. Total turnover has a negative significant effect for firms with a turnover between 500,000 and 10 million euros. The age of the SME has a negative statistically significant effect for firms established between 2010 and 2012. If the firm belongs to the manufacturing activity sector, it has a greater probability to implement this action in the future, the contrary is true for the industrial sector. Workers in green jobs have a significant and positive impact on the probability of undertaking this policy; while the effect is negative if there are no workers with this type of duties. A positive influence of perceiving the need of environmental skills is present for all eight actions. The number of workers in green jobs has a positive effect only if they are at least 10. An important percentage of heterogeneity is estimated: the ICC is equal to 0.2009.

### 3.2.2 Saving energy

63.2% of European SMEs are undertaking measures to save energy and 41.2% plan to implement it as an additional resource efficiency action, as the EU is suggesting. The economic sector has a significant positive effect for manufacturing, negative for industry and services; having at least 10 workers for green jobs has a positive impact. The negative effect of turnover stops at 2 million and starts with a turnover greater than 10 million. For this and the following actions, an increase in turnover corresponds to an increase in the probability of implementation. Heterogeneity between European countries amounts to 27.68%.

### 3.2.3 Using predominantly renewable energy

The European Community Directive on renewable energy (European Community, 2009) requires that at least 20% of Europe's total energy needs are met with renewables by 2020. As can be seen from our data, only 14% of firms had adopted this CE framework in 2017, but 77.6% intend to adopt it. The results of the estimation of the logit regression model show that total turnover has a negative effect on the adoption of this policy when it is very low, below 500,000 euros, but a positive effect when it is very high, over 10 million. No workers in green jobs has a negative effect, while it has a positive effect for firms with between 6 and 50 workers. Heterogeneity among EU MSs is estimated equal to 17.93%, the lowest level of all estimated models.

### 3.2.4 Saving materials

Saving materials is favored by operating in the manufacturing sector, while retail and services have a negative effect; having a yearly turnover lower than 500,000 euros has a negative effect, positive for turnover higher than 50 million. There is no significant effect of the age of the firm. A positive effect is estimated when between 6 and 9 workers are employed in green jobs, negative for 0. This action is undertaken by 56.4% of firms with an ICC equal to 24,21. 50.8% of firms will adopt this additional action.

### 3.2.5 Minimizing waste

Many EU documents (see, for example, European Commission, 2012) refer to the problem of waste reduction. The EU plans for measures to increase waste reuse offer a range of environmental, economic, and social benefits. Our analyses show that only 65.5% of EU firms have adopted this action. This is the most diffused action, however, heterogeneity between countries is the highest, 22.83%. Table 6 reports the inferential results. The probability of implementing this activity increases with the firms' turnover over 10 million - there is negative effect till 500,000 euros. The positive effect of workers in green jobs to the intention to implement the policy in the future starts when they are 100. Age has no effect. ICC equal to 0.2362.

### 3.2.6 Selling scrap material to other companies

In this case, model estimation shows a significant positive effect on the probability of undertaking the action for manufacturing, negative for services; turnover greater than 2 million, being founded before 2010. The effect is negative for retailing and services, the youngest age, yearly turnover lower than 500, 000 euros. The role of green workers on the intention is positive only when they are 100 and that the ICC is 0.1999.

### 3.2.7 Recycling or reusing waste

SMEs operating in the manufacturing sector have a higher probability to implement this action in the future, lower if they offer services. Firm's age has no significant effect while yearly turnover increases the probability only for the highest level, it decreases it when lower than 500,000 euros. As for the other actions, if there are no workers employed in green jobs, the probability decreases, the effect is positive for over 100 employees. This is another action that requires a very big number of workers in green jobs, and consequently big dimension, to be planned for the future. Heterogeneity at country level amounts to 22.94%.

### 3.2.8 Designing sustainable products

By the end of 2017, 1/4 of EU firms are implementing the resource efficiency practice related to designing sustainable products. The positive determinants of this behavior in the future according to the model estimation are firm's turnover over 50 million euros, manufacturing as of activity, more than 10 workers in green jobs. It has a quite large country-level impact (ICC=0.2262).

*Table 5. ML logit model — estimation results – future actions*

| | Actions | Saving water | Saving energy | Renewable energy | Saving materials | Minimizing waste | Selling scrap | Recycling | Products |
|---|---|---|---|---|---|---|---|---|---|
| Intercept | | -0.211* | -0.001 | -0.634* | -0.375* | -0.028 | -0.635* | -0.242* | -0.617* |
| Some | 0.431* | | | | | | | | |
| Many | -0.034* | | | | | | | | |
| Few | 0.087* | | | | | | | | |
| No | -0.483* | | | | | | | | |
| **Activity sector** | | | | | | | | | |
| Manufacturing | -0.149* | 0.097* | 0.120* | 0.052 | 0.252* | 0.214* | 0.466* | 0.155* | 0.311* |
| Retail | 0.029* | 0.037 | 0.015 | -0.020 | -0.148* | -0.039 | -0.056 | -0.039 | -0.164* |
| Services | 0.117* | -0.013 | -0.070* | -0.049 | -0.165* | -0.166* | -0.554* | -0.232* | -0.213* |
| Industry | 0.003 | -0.120* | -0145* | 0.017 | 0.0620 | -0.009 | 0.144* | 0.116* | 0.067 |
| **Date of birth** | | | | | | | | | |
| Before 1 1 2010 | -0.021 | 0.023 | 0.104 | 0.066 | 0.015 | 0.055 | 0.107 | 0.009 | -0.027 |
| 1/1/2010 – 31/12/2012 | 0.0267 | -0.202* | -0.080 | 0.003 | -0.021 | -0.028 | 0.028 | -0.019 | 0.055 |
| 1/1/2013 – 31/12/2017 | -0.024 | -0.045 | 0.046 | 0.080 | 0.106 | 0.118 | 0.147 | 0.050 | 0.049 |
| After 1/1/2017 | 0.019 | 0.225 | -0.071 | -0.149 | -0.100 | -0.146 | -0.283 | -0.041 | -0.077 |
| **Employees in green jobs** | | | | | | | | | |
| 0 | 0.355* | -0.550* | -0.454* | -0.692* | -0.476* | -0.535* | -0.508* | -0.608* | -0.583* |
| 1-5 | 0.004 | -0.024 | 0.082 | -0.074 | 0.022 | 0.003 | -0.029* | 0.011 | 0.086 |
| 6-9 | -0.091 | 0.131 | 0.107 | 0.223* | 0.292* | 0.054 | 0.014 | -0.015 | 0.027 |
| 10-50 | -0.102* | 0.134* | 0.212* | 0.229* | 0.058 | 0.113 | 0.116 | 0.110 | 0.166* |
| 51-100 | -0.012 | -0.002 | -0.062 | 0.105 | 0.023 | 0.034 | 0.002 | 0.038 | 0.017 |
| >100 | -0.153* | 0.313* | 0.116* | 0.209 | 0.081 | 0.331* | 0.404* | 0.464* | 0.288* |
| **Turnover in euros** | | | | | | | | | |
| <100,000 Euros | 0.145* | -0.061 | -0.211* | -0.154* | -0.161* | -0.265* | -0.549* | -0.172* | -0.156 |
| 100,000-500,000 | 0.124* | 0.092* | -0.221* | -0.217* | -0.144* | -0.107* | -0.369* | -0.137* | -0.009 |
| 500,000-2 million | 0.050* | -0.145* | -0.108* | -0.062 | -0.053 | -0.056 | 0.014 | 0.017 | 0.018 |
| 2-10 million | 0.015 | -0130* | -0.037 | 0.085 | -0.0 45 | -0.017 | 0.147* | -0.064 | -0.046 |
| 10-50 million | -0.114* | 0.053 | 0.203* | 0.103 | 0.140* | 0.143* | 0.390* | 0.087 | -0.028 |
| >50 million | -0.220* | 0.375* | 0.374* | 0.225* | 0.263* | 0.302* | 0.367* | 0.270* | 0.221* |
| **Turnover over the past two years** | | | | | | | | | |
| Increased | -0.069* | 0.042 | 0.092* | 0.130* | 0.073* | 0.091* | 0.107* | 0.101* | 0.144* |
| Decreased | 0.058* | -0.020 | -0.066* | -0.088* | -0.089* | -0.114* | -0.096* | -0.118* | -0.117* |
| Unchanged | 0.010 | -0.022 | -0.026 | -0.042 | 0.016 | 0.023 | -0.011 | 0.017 | -0.027 |
| Need of expertise | -0.280* | 0.429* | 0.378* | 0.354* | 0.473* | 0.584* | 0.547* | 0.343* | 0.452* |
| Var($u_j$) | 0.054* | 0.251* | 0.277* | 0.180* | 0.319* | 0.309* | 0.250* | 0.230* | 0.292* |
| ICC | 0.051 | 0.201 | 0.217 | 0.152 | 0.242 | 0.236 | 0.200 | 0.187 | 0.226 |

\* statistically significant effect p-value<0.05

## 4. Concluding remarks

This article provides an overview of the adoption of actions to increase resource efficiency in the European Union, with a specific focus on green employment and environmental skills. Actions related to resource efficiency are still employed by a small number of firms. This is particularly true for small and medium enterprises.

In this paper, we analyzed survey data collected by the European Commission (Flash Eurobarometer 456, European Commission, 2016). Data were collected in September 2017 and the sample was made up of over 10,000 SMEs active in the 28 EU Member States. To ensure that the sample is representative of the entire population, it is comprised of firms with different sizes, ages, and types of activity. The survey data allows to explore the spread of resource efficiency practices and implementation of environmental skills in firms across EU countries and to evaluate the determinants of this behavior, with a specific attention to the characteristics of the SMEs, green employment and eventual homogeneity within countries.

We estimate a multilevel latent class model in order to identify groups of European companies that are homogeneous in their implementation and intention to implement resource efficiency practices and groups of countries that are homogeneous in the composition of companies operating on their territory. Companies are classified in six clusters both for their observed behavior and for intentions. Countries are classified in four classes, but the compositions of these groups differ if we look at undertaken actions and intentions for the future. The analyses show that there is a non-negligible portion of heterogeneity within and between countries and that workers employed in green jobs and a positive attitude towards environmental skills play a role in determining segments' belonging.

These results are confirmed by the estimation of multilevel regression models to explain the number and type of resource efficiency actions in the plans of European SMEs. A general important evidence is that having no workers dedicated to green jobs has a negative effect in all models, while perceiving the need of extra environmental skills has a positive effect on the intention to implement all eight actions in the future. Other characteristics of the firm have a significant impact on resource efficiency, in general, older and bigger firms, with larger yearly turnover, are more prone to implement resource efficiency actions. Type of actions chosen varies across activity sectors.

More than 30% of SMEs in our sample are classified in cluster 6 of Table 3 (no resource efficiency actions implemented), and over 40% of SMEs are classified in cluster 6 of Table 4 (no resources efficiency actions planned for the near future). In order to devise policies that favor resource efficiency implementation, it is important for European countries to know what is the level of implementation of CE practices by SMEs in its territory and also if there are ongoing plans to increase this behavior. SMEs are influenced in this decision by internal and external factors, and the identification of these factors suggests most efficiencient policies to implement. As findings of this paper underline, green employment and green skills are strictly linked to the decision on eco-innovation in SMEs, so that acquisition and diffusion of environmental skills deserves attention and investment at European level.

Moreover, we found that some specific structural characteristics of the SMEs have a significant effect on the decision and the intention to implement CE practices. For example, the economic activity sector reveals its importance. Almost half of European SMEs are active in the service sector, which is the one with the highest percentage of firms not implementing resource efficiency actions, or nor intending to do so; many of these SMEs are located in Italy, Germany, Poland, Spain, France and Greece.

In the reference literature, the relationship between green jobs, environmental skills, workers employed in circular economy practices, and resource efficiency has received limited attention so far.

As this paper shows, there is a significant association among all these factors and there is also an interaction effect with other characteristics of the firms as age, size, turnover, economic activity sector. The aspect the needs more investigation is the direction of this association. The main question is: does the number of workers in green jobs stimulate the adoption resource efficiency practices or as SMEs start to adhere to CE, more workers are employed in green activities? To properly answer to this question, longitudinal data is necessary. However, some results could also be inferred from cross-sectional data. In this work, we look at the question that asks about the intention to implement resource efficiency actions in the near future. In a future research we might exploit data collected over time by Eurobarometer on equivalent - although not the same – samples.

A second topic of further research could explore the potential effect on implementation of resource efficiency action of covariates at country level such as economic and social indicators that might be taken from European official statistics.

Geissdoerfer, M., Savaget, P., Bocken, N.M.P., Hultink, E.J. (2017). The circular economy – a new sustainability paradigm? *Journal of Cleaner Production*, **143**, 757-768.

Ghisellini, P., Cialani, C., Ulgiati, S. (2016). A review on circular economy: the expected transition to a balanced interplay of environmental and economic system. *Journal of Cleaner Production*, **114**, 11-32.

Goodman, L.A. (1974a). The analysis of systems of qualitative variables when some of the variables are unobservable: Part I. A modified latent structure approach. *American Journal of Sociology*, **79**, 1179-1259.

Goodman, L.A. (1974b). Exploratory latent structure analysis using both identifiable and unidentifiable models. *Biometrika*, **61**, 215-231.

Hedeker, D., & Gibbons, R.D. (1994). A random-effects ordinal regression model for multilevel analysis. *Biometrics*, **50**, 933-944.

Horbach, J., & Rammer, C. (2019). Circular economy innovations, growth and employment at the firm level: empirical evidence from Germany. *Journal of Industrial Ecology*, online.

Hox, J. 2002. *Multilevel Analysis: Techniques and Applications. Quantitative Methodology Series*. Mahwah, NJ, US: Lawrence Erlbaum Associates Publishers.

Hughes, R. (2017). The EU Circular Economy package - life cycle thinking to life cycle law?, 24th CIRP Conference on Life Cycle Engineering, *Procedia CIRP*, **61**, 10-16.

Jové-Llopis, E., & Segarra-Blasco, A. (2018). Eco-efficiency actions and firm growth in European SMEs. *Sustainability*, **10**, 281.

Kirchherr, J., Reike, D., & Hekkert, M. (2017). Conceptualizing the circular economy: An analysis of 114 definitions. *Resources, Conservation & Recycling*, **127**, 221-232.

Kirchherr, J., Piscitelli, L., Bour, R., Kostense-Smit, E., Muller, J., Hiubrechtse-Truijens, A., & Kehkkert, M. (2018). Barriers to the Cirular Economy: Evidence from the European Union (EU). *Ecological Economics*, **150**, 264-272.

Korhonen, J., Honkasalo, A., & Seppala, J. (2018). Circular economy: The concept and its limitations. *Ecological Economics*, **143**, 37-46.

Lazarsfeld, P.F. & Henry, N.W. (1968). *Latent structure analysis*. Boston: Houghton Mifflin.

Leoncini, R., Marzucchi, A., Montresor, S., Rentocchini, F., & Rizzo, U. (2019). 'Better late than never': the interplay between green technology and age for firm growth. *Small Business Economics*, **52,** 891–904

Lieder, M., & Rashid, A. (2016). Towards circular economy implementation: A comprehensive review in context of manufacturing industry. *Journal of Cleaner Production*, **115**, 36-51.
23